\documentclass[aps,prl,10pt,twocolumn,preprintnumbers,superscriptaddress,nofootinbib,letterpaper,longbibliography,twoside]{revtex4-1}

\usepackage{amsmath,amssymb} 
\usepackage{graphicx} 
\usepackage[export]{adjustbox} 
\usepackage{url} 
\usepackage[dvipsnames]{xcolor} 
\usepackage{siunitx} 

\usepackage{titlesec}
\usepackage{microtype}
\titlelabel{\thetitle\quad}  

\usepackage{hyperref} 
\usepackage[capitalize]{cleveref} 
\newcommand\myshade{80} 
\colorlet{mylinkcolor}{ForestGreen}
\colorlet{mycitecolor}{Red}
\colorlet{myurlcolor}{violet}
\hypersetup{
  linkcolor  = mylinkcolor!\myshade!black,
  citecolor  = mycitecolor!\myshade!black,
  urlcolor   = myurlcolor!\myshade!black,
  colorlinks = true
}

\newcommand{\github}[1]{\href{https://github.com/#1}{\includegraphics[width=8pt]{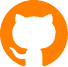}}}

\newcommand{\orcid}[1]{\begingroup
  \hypersetup{hidelinks}\href{https://orcid.org/#1}{\includegraphics[width=10pt]{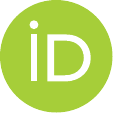}} \endgroup}

\makeatletter
\providecommand*{\diff}%
	{\@ifnextchar^{\DIfF}{\DIfF^{}}}
\def\DIfF^#1{%
	\mathop{\mathrm{\mathstrut d}}%
		\nolimits^{#1}\gobblespace}
\def\gobblespace{%
	\futurelet\diffarg\opspace}
\def\opspace{%
	\let\DiffSpace\!%
	\ifx\diffarg(%
		\let\DiffSpace\relax
	\else
		\ifx\diffarg[%
			\let\DiffSpace\relax
		\else
  			\ifx\diffarg\{%
				\let\DiffSpace\relax
			\fi\fi\fi\DiffSpace}

\begin{document}

\title{Detectable MeV Neutrino Signals from Neutron-Star Common-Envelope Systems}

\author{Ivan Esteban \orcid{0000-0001-5265-2404}}
\email{ivan.esteban@ehu.eus}
\affiliation{Department of Physics, University of the Basque Country UPV/EHU, PO Box 644, 48080 Bilbao, Spain}
\affiliation{Center for Cosmology and AstroParticle Physics (CCAPP), Ohio State University, Columbus, OH 43210, USA}
\affiliation{Department of Physics, Ohio State University, Columbus, OH 43210, USA}
\affiliation{Department of Astronomy, Ohio State University, Columbus, OH 43210, USA}

\author{John F.~Beacom \orcid{0000-0002-0005-2631}}
\email{beacom.7@osu.edu}
\affiliation{Center for Cosmology and AstroParticle Physics (CCAPP), Ohio State University, Columbus, OH 43210, USA}
\affiliation{Department of Physics, Ohio State University, Columbus, OH 43210, USA}
\affiliation{Department of Astronomy, Ohio State University, Columbus, OH 43210, USA}

\author{Joachim Kopp \orcid{0000-0003-0600-4996}\,}
\email{jkopp@cern.ch}
\affiliation{Theoretical Physics Department, CERN,
             1211 Geneva 23, Switzerland}
\affiliation{PRISMA Cluster of Excellence \& Mainz Institute for
             Theoretical Physics, \\
             Johannes Gutenberg University, 55099 Mainz, Germany}

\date{October 30, 2023}

\preprint{MITP-23-062} 


\begin{abstract}
\noindent
Common-envelope evolution --- where a star is engulfed by a companion --- is a critical but poorly understood step in, e.g., the formation pathways for gravitational-wave sources.  However, it has been extremely challenging to identify observable signatures of such systems.  We show that for systems involving a neutron star, the hypothesized super-Eddington accretion onto the neutron star produces MeV-range, months-long neutrino signals within reach of present and planned detectors.  While there are substantial uncertainties on the rate of such events (0.01--1/century in the Milky Way) and the neutrino luminosity (which may be less than the accretion power), this signal can only be found if dedicated new analyses are developed.  If detected, the neutrino signal would probe super-Eddington accretion, leading to significant new insights into the astrophysics of common-envelope evolution.\github{ivan-esteban-phys/common-envelope-thermal}
\end{abstract}


\maketitle


What happens when an ordinary star engulfs a neutron star?  While this sounds exotic, it is believed to happen relatively frequently (0.01--1/century in the Milky Way) in binary and higher-multiplicity star systems, and is an especially interesting example of common-envelope evolution~\cite{%
  Paczynski:1976,      
  Han:1995,            
  Drout:2023atn,       
  Belczynski:2016,     
  Fragos:2019box,      
  Ginat:2019qvu,       
  Hutilukejiang:2018, 
  Ivanova:2013,        
  Ivanova:2020}.        
For instance, it is considered necessary to explain the existence of X-ray binaries and gravitational-wave sources involving neutron stars. Both types of objects require extremely tight binary systems that may be produced through common-envelope evolution. But the process itself is mysterious because it is very difficult to tell if a star is harboring a neutron star.

\begin{figure}[b]
    \centering
    \includegraphics[width=\columnwidth]{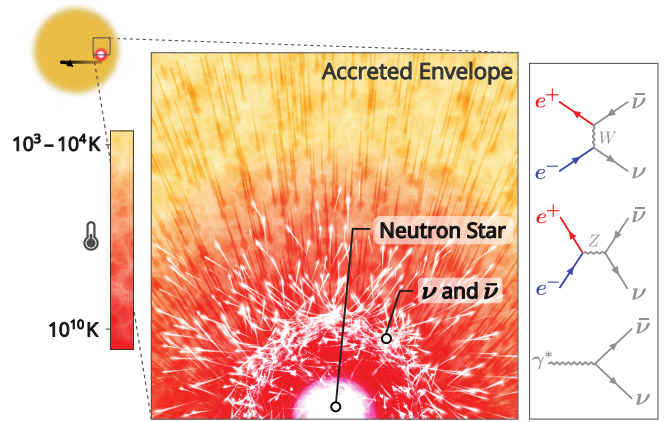}
    \vspace{-0.4cm}
    \caption{Neutrino production via super-Eddington accretion onto a neutron star in a common-envelope system. \emph{Infalling matter releases gravitational potential energy and heats up until cooling via neutrino emission becomes dominant.}}
    \label{fig:cartoon}
\end{figure}

Inside a host star, a neutron star is expected to accrete material at a high rate due to its strong gravitational field and the large ambient density. The process ends when the neutron star merges with the core of the host star or when the envelope is blown off. However, the details are poorly understood~\cite{Ivanova:2013, Ivanova:2020}. If outward radiation pressure counterbalances gravity, accretion is stabilized at the so-called Eddington limit, and gravitational energy is converted into photons. However, since the early 1990s it has been argued that accretion may become so fast that radiation is trapped and pulled inward, leading to super-Eddington (hypercritical) accretion rates~\cite{%
  Chevalier:1989,   
  Houck:1991,       
  Chevalier:1993,   
  Ricker:2012,      
  Ivanova:2013,     
  MacLeod:2014yda,  
  MacLeod:2017,     
  Ivanova:2020}.    
In this case, the matter in the accretion flow would heat up to the point where \emph{neutrino emission becomes the dominant cooling channel}, as illustrated in \cref{fig:cartoon}.

In this \textit{Letter}, we investigate this neutrino emission.  We show that months-long super-Eddington accretion rates of up to $\dot{M} \simeq \SI{0.1}{M_\odot/yr}$, as predicted by several simulations~\cite{%
  Houck:1991,       
  Chevalier:1993,   
  Ricker:2012,      
  Ivanova:2013,     
  MacLeod:2014yda,  
  MacLeod:2017,     
  Ivanova:2020},    
imply a flux of MeV neutrinos that, for a system in the Milky Way, is observable in current and upcoming detectors like Super-Kamiokande, JUNO, and DUNE.  Discovery would require dedicated new analyses, as well as some amount of luck given the low anticipated rate of galactic common-envelope events.  However, without a search, we will definitely miss a neutrino signal that would directly answer many of the most important questions surrounding common-envelope evolution, such as the accretion rate, duration, and outcome. Once a discovery has been made, the key observables will be the neutrino energy, flux, and time profile. For close sources, crude directionality will also be possible.

In the following, we describe the astrophysics of neutron star common-envelope systems; then calculate the neutrino flux, spectrum, and detection prospects; and, last, conclude and discuss directions for future work. In the Supplemental Material, which includes Refs.~\cite{Shapiro:1973, Cox_Giuli, Bildsten:1998km, Egawa:1977, Paxton2011, Paxton2013, Paxton2015, Paxton2018, Paxton2019, Jermyn2023, Rogers2002, Saumon1995, Irwin2004, Timmes2000, Potekhin2010, Itoh:1996, Clark:2022tli, MacLeod:2015, Landau:1959, Kato:2015faa, Dighe:1999bi, Kuo:1989qe, Maltoni:2015kca, Esteban:2020cvm, Brdar:2023ttb, Vogel:1999zy, Strumia:2003zx, Super-Kamiokande:2016yck, JUNO:2023vyz, JUNO:2015zny, Formaggio:2012cpf, Castiglioni:2020tsu, Janka:2012sb}, we provide technical details on our simulations and results.

\textbf{Astrophysical framework.---}\,%
%
We review common-envelope evolution with a neutron star (see also, e.g., Refs.~\cite{Iben:1993yi, Ivanova:2020, Roepke:2022icg}). The process starts when the host star overfills its Roche lobe, either because it expands or because the orbital separation decreases. The neutron star inspirals within the now-common envelope due to drag, accreting material at potentially high rates due to the large ambient density.  The duration of this phase is limited to a time $\sim \mathrm{month} \times (\SI{0.1}{M_\odot/yr}/\dot{M})$~\cite{Iben:1993yi, MacLeod:2014yda}, and possible outcomes include envelope ejection or core merger (forming a Thorne--\.{Z}ytkov object)~\cite{Thorne:1977, Houck:1991, Fryer:1995dr, Popham:1998ab, Brown:1999dda, Narayan:2001, Lee:2005se, Lee:2005et}. This can lead to merger-triggered supernovae~\cite{Chevalier:2012ba, Schroder2020, Soker2017}, including a recent signal that may be explained by common envelope~\cite{Dong2021}.

If the mass accretion rate onto the neutron star is low, gravitational potential energy is primarily released as electromagnetic radiation. In this regime, accretion cannot exceed the Eddington rate, $\sim \SI{e-8}{M_\odot/yr}$ for a neutron star~\cite{Houck:1991}, as radiation pressure hinders further accretion.  However, stable super-Eddington accretion is possible if the inflow velocity is large enough to trap radiation. The primary energy loss process is then thermal neutrino emission~\cite{Basko:1976xn, Houck:1991, Chevalier:1993}. This is thought to occur for accretion rates above $\sim \SI{e-4}{M_\odot/yr}$, though there are significant uncertainties~\cite{Ricker:2012, Ivanova:2013, Fragos:2019box}, partly due to numerical challenges~\cite{Roepke:2022icg} that are shared with other systems such as accretion in Ultra-Luminous X-ray sources~\cite{Basko:1976xn, Mushtukov:2015zea, Suleimanov:2020lum, Chashkina:2019qxu, Brice:2021urc}. One uncertainty regards the need to dissipate angular momentum, but super-Eddington rates as large as $\SI{0.1}{M_\odot/yr}$ are found in simulations both neglecting~\cite{Houck:1991, Brown:1995} and including this effect~\cite{Brown:1999dda, Ricker:2012, MacLeod:2014yda, MacLeod:2017, Hutchinson-Smith:2023apu}. Another is whether all energy is dissipated by neutrino emission or whether other mechanisms such as material ejection in jets, photon convection, or magnetic field-induced opacity decrease play an important role~\cite{MorenoMendez:2017xmw, Murguia-Berthier:2017, Chamandy:2018, Lopez-Camara:2018cym, Lopez-Camara:2020pot, MorenoMendez:2022obo, Law-Smith:2020jwf, 1992ApJ...388..561A, ZhangLiZhongZhang:2022bse, ZhangLiZhong:2021bdw, Turner:2005tx, Basko:1976xn, Mushtukov:2022leu, Basko:1976xn, Mushtukov:2015zea, Suleimanov:2020lum, Chashkina:2019qxu, Brice:2021urc}. The observed neutron star mass distribution, while compatible with super-Eddington accretion, also sets limits~\cite{MacLeod:2014yda, Hutchinson-Smith:2023apu, Ozel:2012ax}. Moreover, transitioning from sub-Eddington ($\lesssim 10^{-8} M_\odot/\mathrm{yr}$) to super-Eddington ($\gtrsim 10^{-4} M_\odot/\mathrm{yr}$) rates requires departures from steady-state spherical accretion, with the details being uncertain~\cite{Chevalier:1993, Houck:1991, Sakurai:2016hns, Sugimura:2016dmy}\mbox{. Direct observational tests are needed.}

The rate of neutron-star common-envelope events in the Milky Way is uncertain. An upper bound is the supernova rate, roughly $1$/century~\cite{Rozwadowska:2020nab}, which limits the neutron-star birth rate; while a lower bound is the theoretically estimated formation rate of Thorne--\.{Z}ytkov objects, roughly $0.01$/century~\cite{Hutilukejiang:2018, Podsiadlowski:1995}. Refs.~\cite{Hutilukejiang:2018, Ginat:2019qvu} estimate a neutron-star common-envelope rate of roughly $0.1$/century.

\textbf{Neutrino flux.---}
%
We first make a simple estimate of the neutrino emission rate for super-Eddington accretion, then describe detailed calculations that support it.

\begin{figure}[t]
    \centering
    \includegraphics[width=1.01\columnwidth]{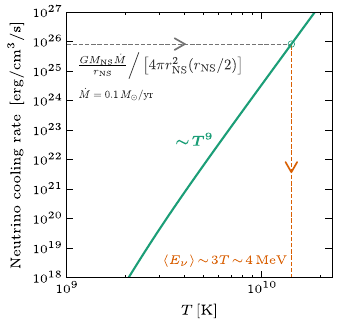}
    \caption{Energy loss due to neutrino emission as a function of temperature (solid). \emph{Due to the steep dependence, the average neutrino energy is robustly determined from the gravitational energy release rate due to accretion} (horizontal dashed).}
    \label{fig:E-loss-rate}
\end{figure}

\Cref{fig:E-loss-rate} shows that the average neutrino energy, $\langle E_\nu \rangle$, is large enough to be detectable. It is similar to the energies of the electrons, positrons, and plasmons from which neutrinos are dominantly produced~\cite{Itoh:1996}, $\langle E_\nu \rangle \sim 3 T \sim \SI{4}{MeV}$, with $T$ the temperature of the accretion flow. $T$ is simply but robustly determined by equating the rate of gravitational potential energy release to the neutrino cooling rate. For the gravitational energy release, we assume $\dot{M} = \SI{0.1}{M_\odot/yr}$, a neutron star of mass $M_\text{NS} = \SI{1.3}{M_\odot}$ and radius $r_\text{NS} = \SI{12}{km}$. Moreover, we assume that the energy is released in a shell of thickness $r_\text{NS}/2$. We follow Refs.~\cite{Dicus:1972yr, Braaten:1993jw} to compute the neutrino cooling rate, which scales $\propto T^9$. Therefore, the specifics of the accretion process hardly matter for determining $\langle E_\nu \rangle$.

The overall neutrino flux scales linearly with the accretion rate. For the above parameters, accretion releases $G M_\mathrm{NS} \dot{M} / r_\mathrm{NS} \simeq \SI{e45}{erg/s}$ of gravitational energy. When carried by neutrinos with energies $\sim \SI{4}{MeV}$, this gives a neutrino production rate $\sim \SI{e50}{s^{-1}}$ or, at a distance of \SI{10}{kpc}, a flux $\sim \SI{e4}{cm^{-2}.s^{-1}}$. Folded with the cross section for inverse beta decay at the relevant energies, $\sim \SI{e-42}{cm^2}$, this leads to $\sim 100$ events over a few months in a detector like Super-Kamiokande.

\begin{figure*}
    \centering
    \includegraphics[height=0.209\textheight, valign=t]{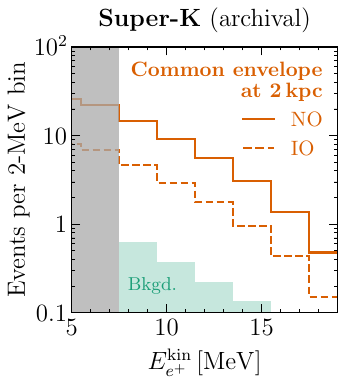}
    \includegraphics[height=0.209\textheight, valign=t]{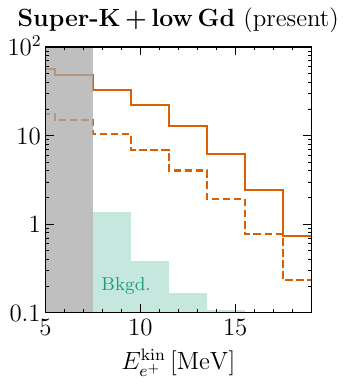}
    \includegraphics[height=0.209\textheight, valign=t]{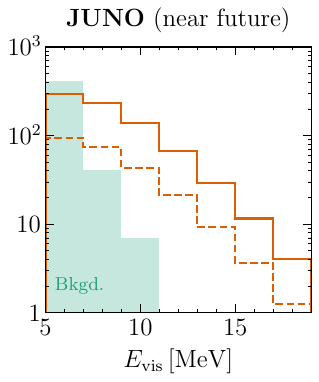}
    \includegraphics[height=0.209\textheight, valign=t]{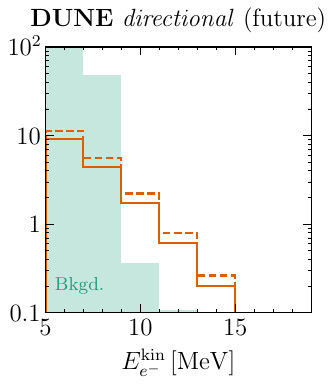}
    \caption{Neutrino signal at different detectors from a super-Eddington neutron-star common-envelope event, integrating over three months (see text for details). Using the astrophysical model from Ref.~\cite{MacLeod:2014yda}, we show the prediction for the normal (NO, solid) and inverted (IO, dashed) mass orderings. Gray regions are below the energy ranges of publicly available background estimates. Super-Kamiokande and JUNO (note the different vertical scale) would detect $\bar{\nu}_e$ with high statistics, whereas DUNE could detect neutrino--electron scattering and thereby roughly localize the source. \emph{A common-envelope neutrino signal can be detectable in past, present, and future detectors.}}
    \label{fig:rates}
\end{figure*}

To go beyond these estimates, we compute the temperature and density profile of the accretion flow by solving the equations of hydrodynamics in a spherical setup, following Ref.~\cite{Houck:1991}. The accretion rate as a function of time is taken from the simulation in Ref.~\cite{MacLeod:2014yda}, which predicts $\dot{M} \sim \SI{e-3}{M_\odot/yr}$ at the onset, increasing to $\sim \SI{0.1}{M_\odot/yr}$ when the neutron star reaches the core of the host after a fraction of a year. We focus on the final three months, during which $\gtrsim 90\%$ of neutrinos are emitted. For a given temperature and density, we compute the neutrino flux from electron--positron annihilation following Ref.~\cite{Dicus:1972yr}, and the subdominant flux from plasmon decay following Ref.~\cite{Braaten:1993jw}. Both processes produce neutrinos of all flavors, with a preference for $\nu_e$ and $\bar\nu_e$. We propagate neutrinos out of the star assuming adiabatic flavor conversion (non-adiabaticities lead to $\lesssim 10\%$ corrections). We also account for gravitational redshift and neutrino absorption by the neutron star~\cite{Fattoyev:2017ybi}. The details of the calculation are given in the Supplemental Material. To support further work, our code is publicly available.\github{ivan-esteban-phys/common-envelope-thermal}

The flavor composition of the emitted neutrino flux depends on the neutrino mass ordering. For the normal ordering, both MSW resonances lie in the neutrino sector. Considering neutrinos and anti-neutrinos that are initially of electron type (the most abundantly produced and easiest to detect flavor), an initial $\nu_e$ will leave the star approximately as a $\nu_3$ mass eigenstate, with only a $\sim 2\%$ $\nu_e$ admixture; and an initial $\bar\nu_e$ will leave as a $\bar\nu_1$, with a $\sim 70\%$ $\bar\nu_e$ component. For the inverted ordering, in contrast, there are resonances for neutrinos and antineutrinos. $\bar\nu_e$ are almost completely converted to other flavors, while $\nu_e$ retain a large electron flavor component.

\textbf{Neutrino detection.---}
%
The most promising detection channels are inverse beta decay ($\bar\nu_e + p \to e^+ + n$, with a large cross section) and neutrino--electron scattering ($\nu + e^{-} \to \nu + e^{-}$, with some directionality). The main backgrounds are due to spallation from cosmic-ray muons~\cite{Li:2014sea, Li:2015lxa, Li:2015kpa, Capozzi:2018dat, Zhu:2018rwc, JUNO:2021vlw}, which produces radioactive isotopes, and neutral-current atmospheric neutrino interactions~\cite{Ankowski:2011ei, T2K:2014vog, Super-Kamiokande:2019hga, T2K:2019zqh, JUNO:2021vlw}. 

We consider the following detectors:
\begin{itemize}

    \item \emph{Super-K (archival).} A new analysis searching for a common-envelope signal using many years of existing Super-Kamiokande data will resemble a search for the diffuse supernova neutrino background~\cite{Beacom:2010kk, Super-Kamiokande:2021jaq}, augmented by criteria exploiting the transient nature of the signal. It would benefit from a well-understood detector, though the detection efficiency would be relatively low (5--15\%) due to the challenges associated with neutron tagging.

    \item \emph{Super-K (low Gd).} The Super-Kamiokande collaboration has recently added a small amount (0.01\% mass fraction) of gadolinium to their detector to improve the tagging of neutrons produced in inverse beta decay~\cite{Beacom:2003nk, Super-Kamiokande:2023xup}. Still, the detection efficiency remains low, 15--30\%.

    \item \emph{Super-K (Gd, optim.),} an optimistic future upgrade of Super-Kamiokande, with enough gadolinium to increase the detection efficiency to 75\% while maintaining the same background levels as in Ref.~\cite{Super-Kamiokande:2023xup}.

    \item \emph{JUNO,} a detector concluding construction with a fiducial volume comparable to that of Super-Kamiokande. Filled with liquid scintillator, JUNO's detection efficiency is large, 73\%~\cite{JUNO:2021vlw}. The detector suffers, however, from sizable reactor and spallation backgrounds due to its location.
    
    \item \emph{Hyper-K,} the successor to Super-Kamiokande with an eight times larger fiducial volume. We assume the same detection efficiency as for \emph{Super-K (archival)}, but larger spallation backgrounds due to the detector's shallower depth~\cite{Hyper-Kamiokande:2018ofw}.

    \item \emph{Hyper-K (Gd),} a hypothetical upgrade of Hyper-Kamiokande with gadolinium loading for the same detection efficiency as \emph{Super-K (Gd, optim.)}.

    \item \emph{DUNE,} a future liquid argon detector that can detect MeV neutrino--electron scattering. The large solar neutrino background in this channel can be removed by exploiting directionality. Solar $\nu_e$ can also scatter on ${}^{40}$Ar, but we first assume that these events can be rejected by identifying final-state gamma rays from nuclear de-excitation~\cite{DUNE:2020ypp} (below, we consider the opposite scenario too).
    
\end{itemize}
Further details on our treatment of these detectors can be found in the Supplemental Material. We, note, however, that  \emph{any} sufficiently large detector with sensitivity to MeV neutrinos --- where data is taken for, e.g., solar or supernova neutrino searches --- is sensitive to common-envelope neutrinos. Yet new analyses are needed to identify a months-long burst. Otherwise, backgrounds accumulate and overcome the common-envelope signal.

\Cref{fig:rates} shows that the neutrino signal from a nearby common-envelope event can be sizable, with the rate well above backgrounds. We use the accretion rates from Ref.~\cite{MacLeod:2014yda} as described above. For models with other values of $\dot{M}$, the spectrum would be very similar, while the normalization would scale $\propto \dot{M}$.

Once a signal has been identified, simple analyses would provide significant astrophysics insight. First, the neutrino spectrum robustly determines the temperature of the accretion flow (see above). Second, since an observable source has to be in the Milky Way (see below), at a distance $\sim$1--10\,kpc, the measured neutrino flux allows estimating the flux at the source. This directly connects to the accretion rate. If there is an electromagnetic counterpart and its distance were found, this estimation would be significantly sharpened. Finally, the duration of the signal can be related to the accretion rate too (see above).

\begin{figure}
    \centering
    \includegraphics[width=\columnwidth]{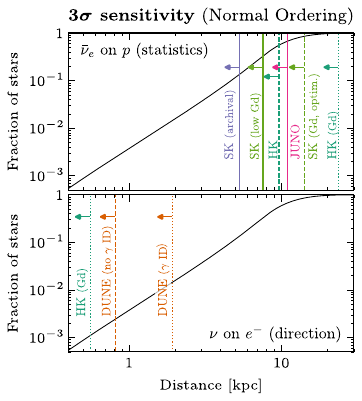}
    \caption{Discovery reach for common-envelope events in the inverse beta decay (top) and neutrino--electron scattering (bottom) channels, using the accretion rates from Ref.~\cite{MacLeod:2014yda}. The black line shows the fraction of Milky Way stars within a given distance from Earth. \emph{Current detectors reach a large fraction of the galaxy, and significant improvement is possible. Directionality is more challenging.}}
    \label{fig:reach}
\end{figure}

\textbf{Discovery reach.---}
%
\Cref{fig:reach} (top panel) shows that current and near-future detectors are sensitive to common-envelope events in a large fraction of the Milky Way, including with archival data, if new analyses are developed. We show the largest distance at which experiments can detect a signal with $3\sigma$ significance, assuming the normal mass ordering (see Supplemental Material for details and for inverted ordering results). The significance does not include a trials factor from all possible 3-month windows. This is appropriate if more than one experiment sees the signal or if there is an electromagnetic counterpart. The impact of the trials factor would be minor, because for sources even slightly closer the significance is much larger as the signal scales with the inverse distance squared. The black line in the plot shows the fraction of Milky Way stars within a given distance from Earth using the thin disk parameters from Refs.~\cite{Girardi:2005fk, Adams:2013ana}, appropriate for common-envelope host stars~\cite{Ivanova:2020}.  For comparison, supernova neutrino bursts can be detected out to hundreds of kpc, which includes the whole Milky Way and its satellites~\cite{Ando:2005ka}. Pre-supernova neutrinos can be detected out to \SI{1}{kpc}~\cite{Li:2020gaz, JUNO:2023dnp, Super-Kamiokande:2022bwp}. Given the galactic supernova rate, this implies that common-envelope signals are more common than pre-supernova signals. \Cref{fig:reach} (bottom panel) shows the same for searches using neutrino--electron scattering. While these are sensitive only to events in our local neighborhood (out to \SI{2}{kpc} at best), they could localize the source in the sky to within about \ang{5}~\cite{DUNE:2020ypp}.

Our assumptions on experimental capabilities are conservative, with much room for improvement. The gadolinium concentration in Super-Kamiokande has already been increased to 0.03\%~\cite{Super-Kamiokande:2023xup} to increase the detection efficiency. The energy cut could also be lowered, as can be seen in \cref{fig:rates}.  If the background due to atmospheric-neutrino neutral-current interactions were greatly reduced, the reach of \emph{Super-K (Gd, optim.)} would change from \SI{14}{kpc} to \SI{21}{kpc}. For both Super-Kamiokande and JUNO, further work on reducing spallation backgrounds~\cite{Li:2015lxa, Super-Kamiokande:2021snn} could increase the reach by up to $\sim$\,20\%. For DUNE, the most important capability is gamma-ray identification, although neutron shielding, as proposed in Refs.~\cite{Capozzi:2018dat, Zhu:2018rwc}, would further increase the reach by $\sim$\,20\%.

\textbf{Outlook and future directions.---}
%
Neutron-star common-envelope events are predicted to occur in the Milky Way at a rate of 0.01--1/century, but it has not been possible to detect them.  Here we show that they would release large MeV-range neutrino fluxes if accretion onto the neutron star is super-Eddington, as suggested by both simple accretion theory and detailed simulations.  Importantly, while the neutrino fluxes are within reach of present and future detectors, finding a signal requires developing dedicated new analyses. 

In short, experiments should look for an excess of MeV-range events with a thermal-like spectrum (we provide the code to compute it \github{ivan-esteban-phys/common-envelope-thermal}) lasting for several months. Experiments sensitive to neutrino--electron scattering can further look for an excess from a fixed position in the sky. While a signal may have already been missed, it will be possible to check archival data.  Our work shows that common-envelope systems are a potential third low-energy astrophysical neutrino source, besides the Sun and supernovae.

Just the fact of a successful detection would reveal both that a neutron star is being engulfed by an ordinary star and that super-Eddington accretion is occurring, resolving important open questions. Simple measurements of the neutrino spectrum and flux would test the conditions at the source. With moderate statistics, the time evolution of the signal would further reveal the dynamics of the inspiraling neutron star.  These would be crucial inputs for more broadly understanding common-envelope evolution, which remains mysterious because the dense overburden of the envelope shields electromagnetic signals.  As an example of a benefit, this would help us understand the pathways that lead to observed gravitational-wave signals from neutron star and black hole mergers.

Identifying the specific star where common envelope occurs remains a long-term challenge. For a nearby event, the signal from neutrino--electron scattering could localize the event to within about five degrees (comparable to supernova burst detection~\cite{Beacom:1998fj}), which would define a search window for an electromagnetic outburst if the envelope is ejected, or even for a supernova~\cite{Chevalier:2012ba, Schroder2020, Soker2017, Dong2021}. To identify a star before that, some breakthrough is needed in electromagnetic, high-energy neutrino, or gravitational-wave signal detection, as suggested in Refs.~\cite{Ivanova:2013, Holgado:2017vut, Ginat:2019qvu, Sridhar:2022uis}. An independent observation of the early stages of a common-envelope event would also reduce backgrounds in the neutrino search by narrowing the relevant time window.  Ultimately, if detection in other channels is successful, even the non-observation of neutrinos from an event would be significant, indicating that super-Eddington accretion is not occurring and setting upper limits on the accretion rate and temperature of the accretion flow.

One interesting direction of future work is the details of incorporation of accreted material into the neutron star. Nuclear reactions deprotonize the infalling matter~\cite{Keegans:2019gje}, which leads to an extra flux of MeV-scale neutrinos. This flux is lower than the one from cooling: a proton gains of order \SI{100}{MeV} in energy during infall, which requires tens of $\sim \SI{4}{MeV}$ neutrinos to dissipate. Nuclear reactions, on the other hand, yield less than one neutrino per infalling proton. The timescale for these nuclear reactions can also be different. Encouragingly, the neutrino flux from deprotonization could be detectable not only from common-envelope events involving neutron stars, but also from those involving white dwarfs, which are far more common. Other astrophysical environments that produce less intense but more frequent neutrino signals, such as that proposed in Ref.~\cite{Asthana:2023vvk}, also deserve detailed studies. Further explorations are needed.

\textbf{Acknowledgments.---}
%
It is a pleasure to thank Simone Bavera, Tassos Fragos, Iñigo Gonzalez, Matthias Kruckow, Shirley Li, Todd Thompson, and Stephan Meighen-Berger for very useful discussions, and Daniel Dominguez (CERN Design and Visual Identity Team) for preparing \cref{fig:cartoon}.  I.E.\ acknowledges support from Eusko Jaurlaritza (IT1628-22).  J.F.B.\ was supported by National Science Foundation Grant No.\ PHY-2310018.

\bibliographystyle{JHEP}
\bibliography{refs}

\clearpage


\onecolumngrid
\let\oldsection\section
\renewcommand{\section}{%
\setcounter{equation}{0}
\setcounter{figure}{0}
\setcounter{table}{0}
\oldsection%
}

\setcounter{section}{0}
\setcounter{secnumdepth}{2}
\setcounter{page}{1}
\makeatletter
\renewcommand{\theequation}{\thesection\arabic{equation}}
\renewcommand{\thefigure}{\thesection\arabic{figure}}
\renewcommand{\thesection}{\Alph{section}}
\renewcommand{\thesubsection}{\thesection.\arabic{subsection}}
\renewcommand{\thesubsubsection}{\thesubsection.\arabic{subsubsection}}
\renewcommand{\thepage}{S\arabic{page}}
\makeatother

\begin{center}
  \textbf{\large Supplemental Material\\[.2cm] Detectable MeV Neutrino Signals from Common-Envelope Systems}\\[0.2cm]
  Ivan Esteban, John F.~Beacom, and Joachim Kopp
\end{center}

Here we provide technical details of our simulations and results to aid further work. We discuss the temperature and density profiles during super-Eddington accretion in \cref{sec:hydro}, neutrino spectra in \cref{sec:neutrino_emission}, neutrino oscillations in \cref{sec:neutrino_oscillations}, neutrino detection in \cref{sec:experiments}, and the distance reach in the inverted mass ordering in \cref{sec:distance}.

As discussed in the main text, the neutrino flux and average energy are robustly determined by energy conservation. The details discussed in this Supplemental Material allow us to sharpen the predictions, but they are not important for our conclusions.

\section{Hydrodynamic equations}
\label{sec:hydro}

We describe how we compute the temperature and density profile of the accretion flow to obtain the neutrino flux. The procedure largely follows Ref.~\cite{Houck:1991}.

As gas is accreted by the neutron star, its evolution is governed by the equations of hydrodynamics. Ref.~\cite{Chevalier:1989} found that the process can be described by a sequence of steady states. We further assume spherical symmetry~\cite{Houck:1991, Murguia-Berthier:2017}. For the large inflow velocities in super-Eddington accretion, radiation diffusion can be neglected~\cite{Houck:1991} and the equations for steady-state spherically symmetric accretion including relativistic effects read~\cite{Houck:1991, Shapiro:1973}
\begin{alignat}{3}
    \frac{1}{r^2} \frac{\diff (r^2 \rho v)}{\diff r}
        &= 0 \,,
     && \text{(continuity equation)}
         \label{eq:continuity} \\
    \frac{\diff(\rho c^2 + e)}{\diff r} - \frac{w}{\rho} \frac{\diff \rho}{\diff r}
    &= \frac{\varepsilon_\mathrm{nuc} - \mathcal{L}_\nu}{v}
        \,,
    \qquad   &&\text{(energy conservation equation)}
        \label{eq:energy} \\
    v \frac{\diff v}{\diff r} + \frac{G M_\mathrm{NS}}{r^2} + \frac{1}{w}\frac{\diff P}{\diff r}\left(v^2 + c^2 - \frac{2 G M_\mathrm{NS}}{r} \right) &= 0 \,,
    &&\text{(Euler equation)}
        \label{eq:Euler} 
\end{alignat}
with $r$ the radial coordinate, $\rho$ the mass density, $v$ the radial component of the fluid four-velocity, $G$ Newton's constant, $M_\mathrm{NS}$ the neutron star mass, $P$ the fluid pressure, $w \equiv \rho c^2 + e + P$ the relativistic enthalpy, with $e$ the fluid internal energy density and $c$ the speed of light, $\varepsilon_\mathrm{nuc}$ the nuclear energy generation rate per unit volume, and $\mathcal{L}_\nu$ the energy loss rate per unit volume due to neutrinos. Here $\rho$, $e$, $P$, $\varepsilon_\mathrm{nuc}$, and $\mathcal{L}_\nu$ are evaluated in the comoving fluid frame. 

\Cref{eq:continuity} can be integrated to obtain $\rho$ as a function of $r$, $v$, and the accretion rate $\dot{M}$. Using Gauss's theorem,
\begin{equation}
\rho(r) = -\frac{\dot M}{4\pi r^2 v(r)}\, .
\label{eq:continuity_2}
\end{equation}
Physically, this equation represents compression due to spherically-symmetric accretion and mass conservation. Note that ${v(r) < 0}$ because the fluid is accreted inwards.

To simplify \cref{eq:energy}, we write $e$ as a function of temperature and density, $e(r) = e(T(r), \rho(r))$. Using standard thermodynamic relations~\cite{Cox_Giuli} and \cref{eq:continuity}, energy conservation can then be written as~\cite{Jermyn2023}
\begin{equation}
\frac{\diff T}{\diff r} = \frac{\varepsilon_\mathrm{nuc} - \mathcal{L}_\nu}{\rho c_V v} - T(\Gamma_3-1)\left(\frac{2}{r} + \frac{1}{v}\frac{\diff v}{\diff r} \right)\, ,
\label{eq:energy_2}
\end{equation}
with 
\begin{equation}
    c_V \equiv \left.\frac{\partial (e/\rho)}{\partial T}\right|_\rho
\end{equation}
the constant-density specific heat per unit mass and 
\begin{equation}
    \Gamma_3 \equiv 1 + \frac{\rho}{T}\left.\frac{\partial T}{\partial \rho}\right|_s
\end{equation}
the adiabatic index, with $s$ entropy per unit mass. The first term on the right-hand side of \cref{eq:energy_2} represents heating due to nuclear reactions and cooling due to neutrino emission, and the second term represents compressional heating.

Similarly, we can simplify \cref{eq:Euler} by writing both $P$ and $e$ as functions of temperature and density, i.e., ${P(r) = P(T(r), \rho(r))}$ and $e(r) = e(T(r), \rho(r))$. Using again standard thermodynamic relations~\cite{Cox_Giuli} and \cref{eq:continuity}, the Euler equation becomes
\begin{equation}
v\frac{\diff v}{\diff r} = -\frac{G M_\mathrm{NS}}{r^2} + \left[ c_s^2\left(\frac{2}{r} + \frac{1}{v} \frac{\diff v}{\diff r} \right) - \frac{\Gamma_3 - 1}{v w/c^2}(\varepsilon_\mathrm{nuc} - \mathcal{L}_\nu) \right] \left( \frac{v^2}{c^2} + 1 - \frac{2 G M_\mathrm{NS}}{r c^2} \right)\, , \label{eq:Euler_2}
\end{equation}
with 
\begin{equation}
c_s^2 \equiv \left.\frac{\partial P}{\partial \rho}\right|_s \frac{1}{1 + (e+P)/(\rho c^2)}
\end{equation}
the sound speed squared accounting for relativity. The first term in \cref{eq:Euler_2} represents acceleration due to gravity, and the second term pressure-induced acceleration.

\subsection{Including nuclear reactions}

Thermonuclear reactions affect the evolution through $\varepsilon_\mathrm{nuc}$ and potentially the equation of state due to a modified composition. However, we find that infalling matter achieves high temperatures ($\sim$ MeV) very rapidly (after $\sim$ 10\,s), which enables efficient photodisintegration and leads the system into nuclear statistical equilibrium where $\varepsilon_\mathrm{nuc} = 0$. 

Nevertheless, at such high temperatures electron capture on a free proton is energetically allowed. (We neglect neutron decay, as material gets incorporated into the neutron star over a timescale $\mathcal{O}(\SI{10}{s})$, much shorter than the neutron lifetime.) Furthermore, because of mass conservation, spherically symmetric accretion significantly increases the density of infalling matter, as parametrized by \cref{eq:continuity_2}. This increases the electron Fermi energy and enables further electron capture. Taking both effects into account, the electron capture rate per free proton is~\cite{Bildsten:1998km}
\begin{equation}
    R_\mathrm{ec} = \frac{\log 2}{ft} \frac{1}{(m_e c^2)^5} \int_Q^\infty \frac{E (E-Q)^2 \sqrt{E^2-m_e^2c^4}}{1 + e^{\frac{E - \mu}{k_B T}}}\,\diff E \, ,
\end{equation}
with $ft = \SI{1065}{s}$, $m_e$ the electron mass, $Q=\SI{1.29}{MeV}$ the energy threshold, $\mu$ the electron chemical potential including the electron mass, $k_B$ Boltzmann's constant, and $\log$ the natural logarithm. Each electron capture consumes \SI{1.29}{MeV} of kinetic energy that go into the proton--neutron mass difference, as well as an energy~\cite{Egawa:1977}
\begin{equation}
    Q_\nu = \left[\int_Q^\infty \frac{E (E-Q)^3 \sqrt{E^2-m_e^2c^4}}{1 + e^{\frac{E - \mu}{k_B T}}}\,\diff E\right] \left/ \left[ \int_Q^\infty \frac{E (E-Q)^2 \sqrt{E^2-m_e^2c^4}}{1 + e^{\frac{E - \mu}{k_B T}}}\,\diff E \right] \right. \ ,
\end{equation}
that gets carried away by neutrinos. This additional cooling channel reduces slightly the average thermal neutrino energy. On top of that, it produces a subleading neutrino signal. In our prediction for the total neutrino signal, we conservatively neglect this component. Its properties depend on the inflow density and the accretion timescale (i.e., on assuming spherical symmetry) and are thus more model-dependent.

To take electron capture into account in the equations of hydrodynamics, we follow the electron fraction
\begin{equation}
    Y_e \equiv \frac{n_{e^-} - n_{e^+}}{n_\mathrm{p} + n_\mathrm{n}} \, ,
\end{equation}
with $n_{e^-}$, $n_{e^+}$, $n_\mathrm{p}$, and $n_\mathrm{n}$ the number densities of electrons, positrons, protons, and neutrons respectively (including nucleons bound inside nuclei). Its steady-state evolution is given in a spherically symmetric system by
\begin{equation}
    \frac{\diff Y_e}{\diff r} = - \frac{R_\mathrm{ec}}{v} \frac{n_\mathrm{H}}{n_\mathrm{p} + n_\mathrm{n}}
    \label{eq:Ye_evolution}
\end{equation}
with $n_\mathrm{H}$ the number density of hydrogen. The energy consumption rate is
\begin{equation}
    \varepsilon_\mathrm{nuc} = - (Q + Q_\nu) R_\mathrm{ec} n_\mathrm{H} \, .
\end{equation}
\Cref{eq:energy_2,eq:Euler_2,eq:Ye_evolution} are three first-order equations with three unknowns, $T(r)$, $v(r)$, and $Y_e(r)$; we solve these equations numerically. We obtain the thermodynamic input quantities $\mathcal{L}_\nu$, $c_V$, $\Gamma_3$, $e$, $P$, $c_s^2$, and $\mu$ as a function of temperature, density, and electron fraction using the equation of state and neutrino cooling modules of \texttt{MESA}~\cite{Paxton2011, Paxton2013, Paxton2015, Paxton2018, Paxton2019, Jermyn2023, Rogers2002, Saumon1995, Irwin2004, Timmes2000, Potekhin2010, Itoh:1996}. As mentioned above, the chemical composition at high temperatures follows nuclear statistical equilibrium, which we compute using the public code \texttt{pynucastro}~\cite{Clark:2022tli}.

\subsection{Accretion shock}

In principle, \cref{eq:energy_2,eq:Euler_2,eq:Ye_evolution}, together with boundary conditions at large $r$, should provide the temperature and velocity profile of the accretion flow. However, to account for the hard surface of the neutron star, we need to impose the additional condition $v(r_\mathrm{NS})=0$, and there is no continuous, spherically-symmetric accretion flow satisfying that condition. This can be seen by rewriting \cref{eq:Euler_2} as
\begin{equation}
     \frac{1}{2} \frac{\diff (v^2)}{\diff r} = \frac{v^2}{v^2 - c_s^2 F_\mathrm{rel}}\left[-\frac{G M_\mathrm{NS}}{r^2} + \frac{2 c_s^2 F_\mathrm{rel}}{r} - \frac{\Gamma_3 - 1}{v w/c^2}(\varepsilon_\mathrm{nuc} - \mathcal{L}_\nu) F_\mathrm{rel}\right]\, , \label{eq:Euler_sonic}
\end{equation}
with $F_\mathrm{rel} \equiv \left(v^2/c^2 + 1 - \frac{2GM_\mathrm{NS}}{r c^2}\right)$ the relativistic correction factor. In the outermost layers of the accretion flow, material is accelerating towards the neutron star, so $\diff (v^2) / \diff r < 0$. Accretion happens inside the Bondi radius ${r_\mathrm{Bondi} \equiv \frac{G M_\mathrm{NS}}{2 c_s^2}}$~\cite{Houck:1991, MacLeod:2015}, which implies that the term in square brackets is negative (nuclear energy generation and neutrino cooling are only relevant very close to the neutron star), and thus $v^2 > c_s^2$. Since $\diff (v^2) / \diff r < 0$, both conditions are always satisfied as $r$ decreases, the right-hand side of \cref{eq:Euler_sonic} is always negative, and there is no continuous solution where $\diff (v^2) / \diff r$ changes sign, as required to decelerate and match with $v=0$ at the neutron star surface. The solution requires a discontinuous accretion shock. The shock jump conditions are~\cite{Houck:1991, Landau:1959}
\begin{align}
    \rho_2 v_2 & = \rho_1 v_1 \label{eq:shock_1}\\
    w_2 v_2 \sqrt{v_2^2 + c^2} & = w_1 v_1 \sqrt{v_1^2 + c^2} \label{eq:shock_2}\\
    w_2 v_2^2/c^2 + P_2 & = w_1 v_1^2/c^2 + P_1 \label{eq:shock_3}
\end{align}
where the subscript 1 refers to quantities before the shock and the subscript 2 to quantities after it. \Cref{eq:shock_1} represents the conservation of mass, and \cref{eq:shock_2,eq:shock_3} represent the conservation of $T^{0r}$ and $T^{rr}$, respectively, with $T^{\mu \nu}$ the fluid stress-energy tensor. We neglect general relativity corrections, as the shock happens far from the neutron star surface. Simple solutions exist if $c_s \ll v \ll c$ before the shock and the gas has a polytropic equation of state~\cite{Landau:1959},
\begin{align} 
    v_2 &= v_1 \frac{\gamma - 1}{\gamma + 1} \, , \label{eq:postshock_1}\\
    \rho_2 &= \rho_1 \frac{\gamma + 1}{\gamma - 1} \, , \label{eq:postshock_2}\\
    P_2 &= \frac{2}{\gamma + 1}\rho_1 v_1^2 \, , \label{eq:postshock_3}
\end{align}
with $\gamma$ the polytropic index. The properties after the shock depend only on the velocity and density before the shock (linked via \cref{eq:continuity_2}), which we have checked to hold in our full simulation to a good approximation. Physically, the shock converts the kinetic energy of the ingoing fluid (accelerated due to the strong gravitational field of the neutron star) into heat.

To integrate \cref{eq:energy_2,eq:Euler_2,eq:Ye_evolution} together with an accretion shock, we proceed as follows
\begin{enumerate}
    \item We set boundary conditions at a large radius $r_\infty$: a chemical composition like that of the Sun, a free-fall velocity ${v(r_\infty) = -\sqrt{2 G M_\mathrm{NS}/r_\infty}}$ (if $v(r_\infty)$ is different, the solution quickly converges to free fall), and a temperature $T(r_\infty)=\SI{5000}{K}$ representative of a red giant envelope (the properties after the shock, where neutrino emission happens, are largely insensitive to $T(r_\infty)$ as described above).
    \item We solve \cref{eq:energy_2,eq:Euler_2,eq:Ye_evolution}, obtaining a pre-shock velocity, temperature, and composition profile that, as described above, does not satisfy $v(r_\mathrm{NS}) = 0$.
    \item We set the shock radius to an arbitrary value $r_\mathrm{shock}$. Starting with the pre-shock profile, we solve \cref{eq:shock_1,eq:shock_2,eq:shock_3} to obtain the post-shock variables. Setting those variables as boundary conditions at $r=r_\mathrm{shock}$, we solve \cref{eq:energy_2,eq:Euler_2,eq:Ye_evolution} down to $r=r_\mathrm{NS}$.
    \item We vary $r_\mathrm{shock}$ until $v(r_\mathrm{NS}) = 0$.
\end{enumerate}

\subsection{Results}

\begin{figure}[hbtp]
    \centering
    \includegraphics[height=0.29\textwidth]{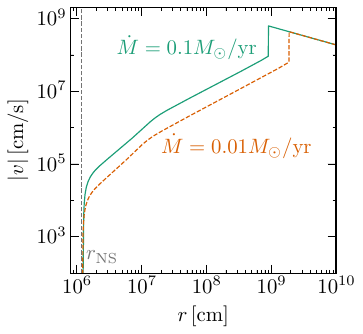}     \includegraphics[height=0.29\textwidth]{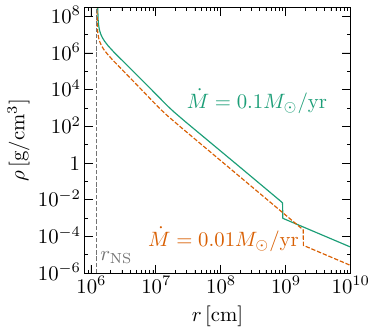}     \includegraphics[height=0.29\textwidth]{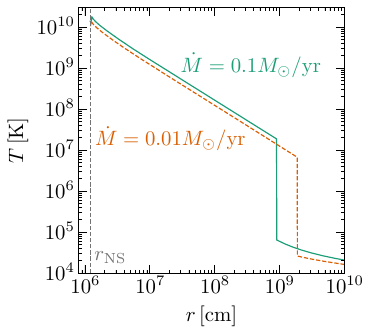}
    \caption{Velocity, density, and temperature profiles for Super-Eddington accretion flow onto a neutron star with a mass of \SI{1.33}{M_\odot} and a radius of \SI{12}{km}. We show the results for accretion rates of \SI{0.1}{M_\odot/yr} and \SI{0.01}{M_\odot/yr}, respectively.}
    \label{fig:accretion_profile}
\end{figure}

\Cref{fig:accretion_profile} shows the velocity, density, and temperature profiles for two values of super-Eddington accretion rates. As in the main text, we adopt median neutron star parameters, $M_\mathrm{NS} = \SI{1.3}{M_\odot}$ and $r_\mathrm{NS}=\SI{12}{km}$. Neutrino emission happens in the innermost region, where $T \gtrsim \SI{e10}{K}$. As described in the main text, the temperature where neutrino emission happens (that determines the neutrino average energy) is only weakly sensitive to the accretion rate, ${T \propto \dot{M}^{1/9}}$: even if $\dot{M}$ gets reduced by an order of magnitude, $T$ only gets reduced by about $30\%$. We find good agreement with the results in Ref.~\cite{Houck:1991}.

\section{Neutrino emission}
\label{sec:neutrino_emission}

We describe how we compute the neutrino flux given the temperature and density profiles. We follow Ref.~\cite{Dicus:1972yr} to compute the flux from electron--positron annihilation, and Ref.~\cite{Braaten:1993jw} to compute the flux from plasmon decay. We find that, due to the high temperature of the accretion flow, positron annihilation produces a much larger neutrino flux than plasmon decay. In this section, we use natural units ($\hbar = c = k_B = 1$).

\subsection{Positron annihilation}

Thermally produced positrons can annihilate with electrons to produce $\nu \bar\nu$ pairs. The number of reactions per unit volume per unit time is~\cite{Dicus:1972yr}
\begin{equation}
    R_{e^+e^-} = \int\!\frac{\diff^3 {\vec p}_\nu}{(2\pi)^3 2E_\nu} \frac{\diff^3 {\vec p}_{\bar\nu}}{(2\pi)^3 2E_{\bar\nu}} \frac{\diff^3 {\vec p}_{e^-}}{(2\pi)^3 2 E_{e^-}} \frac{\diff^3 {\vec p}_{e^+}}{(2\pi)^3 2 E_{e^+}} \, (2\pi)^4 \delta^4(p_{e^-}+p_{e^+}-p_\nu-p_{\bar\nu}) \, f_{e^-}(E_{e^-}) \, f_{e^+} (E_{e^+}) \, |\mathcal{M}|^2\, ,
\end{equation}
with $p_\nu$, $p_{\bar\nu}$, $p_{e^-}$, and $p_{e^+}$ the four-momenta of the neutrino, antineutrino, electron, and positron, respectively; $\vec{p}_\nu$, $\vec{p}_{\bar\nu}$, $\vec{p}_{e^-}$, and $\vec{p}_{e^+}$ their three-momenta; $E_\nu$, $E_{\bar\nu}$, $E_{e^-}$, and $E_{e^+}$ their energies; $f_{e^-}$ and $f_{e^+}$ Fermi-Dirac distributions for electrons and positrons, respectively; and $|\mathcal{M}|^2$ the weak-interaction matrix element
\begin{equation}
    |\mathcal{M}|^2 = 8 G_F^2 \left[(C_V - C_A)^2 (p_{e^-}\cdot p_\nu) (p_{e^+} \cdot p_{\bar\nu}) + (C_V + C_A)^2 (p_{e^-} \cdot p_{\bar\nu})(p_{e^+} \cdot p_\nu) + m_e^2(C_V^2 - C_A^2)(p_\nu \cdot p_{\bar\nu}) \right] \,.
\end{equation}
In this expression, $G_F$ is Fermi's constant, $m_e$ the electron mass, and $C_V$ and $C_A$ the vector and axial couplings that depend on the neutrino flavor,
\begin{alignat}{3}
    C_V & = \frac{1}{2} + 2 \sin^2 \theta_W \qquad \qquad && \text{for $\nu_e$,} \label{eq:CVe} \\
    C_V & = -\frac{1}{2} + 2 \sin^2 \theta_W && \text{for $\nu_\mu$ and $\nu_\tau$,} \label{eq:CVx} \\
    C_A & = \frac{1}{2} && \text{for $\nu_e$,} \label{eq:CAe} \\
    C_A & = -\frac{1}{2} && \text{for $\nu_\mu$ and $\nu_\tau$.} \label{eq:CAx} 
\end{alignat}
Here, $\theta_W$ is the weak mixing angle. Following Ref.~\cite{Kato:2015faa}, we write 
\begin{equation}
    \frac{\diff R_{e^+e^-}}{\diff E_\nu \diff E_{\bar\nu} \diff \cos \theta} = \frac{4 E_\nu E_{\bar \nu}}{(2\pi)^6} G_F^2 \left[\beta_1 I_1 + \beta_2 I_2 + \beta_3 I_3 \right] \, , \label{eq:positron_rate}
\end{equation}
where $\theta$ is the angle between the neutrino and antineutrino, the coefficients $\beta_i$ ($i \in \{1,2,3\}$) are defined as
\begin{align}
    \beta_1 & \equiv (C_V - C_A)^2 \, , \label{eq:beta1}\\
    \beta_2 & \equiv (C_V + C_A)^2 \, , \label{eq:beta2}\\
    \beta_3 & \equiv C_V^2 - C_A^2 \, ; \label{eq:beta3}
\end{align}
and the integrals $I_i$ ($i \in \{1,2,3\}$) are
\begin{align}
    I_1 & = \int \frac{\diff^3 {\vec p}_{e^-}}{(2\pi)^3 2 E_{e^-}} \frac{\diff^3 {\vec p}_{e^+}}{(2\pi)^3 2 E_{e^+}} \delta^4(p_{e^-}+p_{e^+}-p_\nu-p_{\bar\nu}) \frac{f_{e^-}(E_{e^-}) f_{e^+}(E_{e^+})}{4} (p_{e^-}\cdot p_\nu)^2\, , \label{eq:I1} \\
    I_2 & = \int \frac{\diff^3 {\vec p}_{e^-}}{(2\pi)^3 2 E_{e^-}} \frac{\diff^3 {\vec p}_{e^+}}{(2\pi)^3 2 E_{e^+}} \delta^4(p_{e^-}+p_{e^+}-p_\nu-p_{\bar\nu}) \frac{f_{e^-}(E_{e^-}) f_{e^+}(E_{e^+})}{4} (p_{e^-}\cdot p_{\bar\nu})^2\, , \label{eq:I2}\\
    I_3 & = \int \frac{\diff^3 {\vec p}_{e^-}}{(2\pi)^3 2 E_{e^-}} \frac{\diff^3 {\vec p}_{e^+}}{(2\pi)^3 2 E_{e^+}} \delta^4(p_{e^-}+p_{e^+}-p_\nu-p_{\bar\nu}) \frac{f_{e^-}(E_{e^-}) f_{e^+}(E_{e^+})}{4} m_e^2 (p_\nu\cdot p_{\bar\nu})\, .  \label{eq:I3}
\end{align}
We integrate over $\vec{p}_{e^+}$ by writing
\begin{equation}
    \int \frac{\diff^3 {\vec p}_{e^+}}{E_{e^+}} = 2 \int \diff^4 p_{e^+} \delta(p_{e^+}^2 - m_e^2) \Theta(E_{e^+}) \,
\end{equation}
(with $\Theta$ the Heaviside step function), and then eliminating the integral using the $\delta^{(4)}$ functions in \cref{eq:I1,eq:I2,eq:I3}. To carry out the remaining integrals, we take $\vec{p}_\nu + \vec{p}_{\bar\nu}$ along the $z$-axis, and we obtain
\begin{equation}
\begin{split}
    I_1(E_\nu, E_{\bar\nu}, \cos \theta) = -\frac{2\pi T E_\nu^2 E_{\bar\nu}^2(1-\cos \theta)^2}{\Delta_e^5 \left[e^{(E_\nu + E_{\bar\nu})/T} - 1\right]} \lbrace & A T^2 [y_\mathrm{max}^2 G_0(y_\mathrm{max}) - y_\mathrm{min}^2 G_0(y_\mathrm{min}) \\
    &\phantom{A T^2 [}+ 2 y_\mathrm{max} G_1(y_\mathrm{max}) - 2 y_\mathrm{min} G_1(y_\mathrm{min}) \\
    &\phantom{A T^2 [}+G_2(y_\mathrm{max}) - G_2(y_\mathrm{min})] \\
    & + B T[y_\mathrm{max}G_0(y_\mathrm{max})-y_\mathrm{min}G_0(y_\mathrm{min}) \\
    & \phantom{+ B T[}+G_1(y_\mathrm{max}) - G_1(y_\mathrm{min})] \\
    & + C [G_0(y_\mathrm{max}) - G_0(y_\mathrm{min})]\rbrace \, ,
\end{split} 
\end{equation}
where $T$ is the temperature of the system,
\begin{align}
    \Delta_e & \equiv \sqrt{E_\nu^2 + E_{\bar{\nu}}^2+2E_\nu E_{\bar\nu}\cos \theta} \, ,\\
    A & \equiv E_{\bar \nu}^2 + E_\nu^2 - E_\nu E_{\bar \nu} (3 + \cos \theta) \, ,\\
    B & \equiv E_\nu [-2 E_\nu^2 + E_{\bar\nu}^2(1+3\cos \theta) + E_\nu E_{\bar\nu}(3-\cos \theta)] \, ,\\
    C & \equiv E_\nu^2 \left[(E_\nu + E_{\bar\nu}\cos \theta)^2 - \frac{1}{2} E_{\bar\nu}^2 (1 - \cos^2 \theta) - \frac{1}{2E_\nu^2} m_e^2 \Delta_e^2 \frac{1 + \cos \theta}{1 - \cos \theta}\right] \, , \\
    y^\mathrm{max}_\mathrm{min} & \equiv \frac{E_\nu + E_{\bar\nu} \pm \Delta_e \sqrt{1 - \frac{2 m_e^2}{E_\nu E_{\bar\nu}(1-\cos \theta)}}}{2 T} \, , \\
    G_n(y) & \equiv F_n\left(\frac{E_\nu + E_{\bar \nu} + \mu}{T} - y\right) - F_n\left(\frac{\mu}{T}-y\right) \, , \\
    F_n(y) & \equiv \int_0^\infty \frac{t^n}{e^{t-x}+1}\,\diff t \, ,
\end{align}
and $\mu$ is the electron chemical potential including electron mass. The other integrals are simpler,
\begin{align}
    I_2(E_\nu, E_{\bar\nu}, \cos \theta) & = I_1(E_\nu = E_{\bar \nu}, E_{\bar \nu} = E_\nu, \cos \theta) \, , \\
    I_3(E_\nu, E_{\bar\nu}, \cos \theta) & = - \frac{2 \pi T m_e^2 E_\nu E_{\bar \nu} (1-\cos \theta)}{\Delta_e \left[ e^{(E_\nu + E_{\bar\nu})/T} - 1\right]} [ G_0(y_\mathrm{max}) - G_0(y_\mathrm{min})] \, .
\end{align}
The number of neutrinos produced per unit volume, time, and energy is then
\begin{equation}
    \frac{\diff R_{e^+e^-}}{\diff E_\nu} = \int  \frac{\diff R_{e^+e^-}}{\diff E_\nu \diff E_{\bar\nu} \diff \cos \theta} \, \diff E_{\bar \nu} \, \diff \cos \theta \, ,
\end{equation}
where $\cos \theta \in \left[-1, 1-2 m_e^2/(E_\nu E_{\bar\nu})\right]$. The antineutrino spectrum is identical to the neutrino spectrum. 

\begin{figure}[hbtp]
    \centering
    \includegraphics[width=0.5\textwidth]{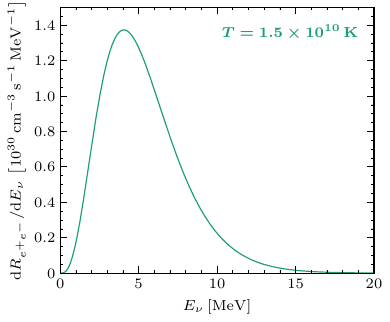}
    \caption{All-flavor neutrino spectrum due to thermal positron annihilation, at temperatures typical in a super-Eddington common-envelope event. The antineutrino spectrum is identical.}
    \label{fig:spectrum_positron}
\end{figure}

\Cref{fig:spectrum_positron} shows the neutrino spectrum at a temperature of \SI{1.5e10}{K}, characteristic of super-Eddington common-envelope events (see \cref{fig:E-loss-rate,fig:accretion_profile}). At such high temperatures, electrons are mainly of thermal origin, so the spectrum is independent of density to a good approximation. As mentioned in the main text, the average neutrino energy is on the order of the average electron energy, and the neutrino spectrum is approximately thermal. We have checked that the energy loss rate, $\displaystyle \int (E_\nu + E_{\bar{\nu}}) \frac{\diff R_{e^+e^-}}{\diff E_\nu \diff E_{\bar\nu} \diff \cos \theta} \diff E_{\nu}\, \diff E_{\bar \nu} \, \diff \cos \theta$, reproduces the results in Refs.~\cite{Dicus:1972yr, Itoh:1996}.

\subsection{Plasmon decay}

We compute the subdominant neutrino flux from plasmon decays closely following Ref.~\cite{Braaten:1993jw}. The neutrino emission rate per unit volume per unit energy receives contributions from transverse and longitudinal plasmon decay,
\begin{align}
    \frac{\diff R_{\gamma^*}}{\diff E_\nu}
                            &= \int\!\frac{4\pi k^2 \diff k}{(2\pi)^3} \, \Big[
                                   2 \, n_B[\omega_t(k)]\, \frac{\diff\Gamma_t(k)}{\diff E_\nu}
                                      + n_B[\omega_l(k)]\, \frac{\diff\Gamma_l(k)}{\diff E_\nu} \Big] \,,
    \label{eq:dNnu-dx-dt-dE}
\end{align}
where the integral runs over plasmon momenta $k$, 
\begin{equation}
    n_B(\omega) = \frac{1}{e^{\omega/T} - 1}
\end{equation}
is the Bose--Einstein distribution describing the plasmon population, and $\omega_t(k)$ and $\omega_l(k)$ are the frequencies of transverse and longitudinal plasmons, respectively. The latter quantities are the solutions of the dispersion relations, which in the Coulomb gauge read
\begin{align}
    [\omega_t(k)]^2 - k^2 - \Pi_t^{(e)}[\omega_t(k), k] - \Pi_t^{(p)}[\omega_t(k), k] &= 0 \,, \label{eq:omega-t} \\[0.2cm]
    [k^2 - \Pi_l^{(e)}[\omega_t(k), k] - \Pi_l^{(p)}[\omega_t(k), k] &= 0 \,. \label{eq:omega-l}
\end{align}
They contain the polarization functions (self-energies)
\begin{align}
    \Pi_t^{(f)}(\omega, k) &= \frac{4 \alpha}{\pi} \int_0^\infty \! \diff p \,
                        \frac{p^2}{E_f} \bigg( \frac{\omega^2}{k^2}
                          - \frac{\omega^2 - k^2}{k^2} \frac{\omega}{2 v_f k} \log \frac{\omega + v_f k}{\omega - v_f k} \bigg)
                        \big[ n_F(E_f) + \bar{n}_F(E_f) \big]  \,,
                                    \label{eq:Pi-t} \\
    \Pi_l^{(f)}(\omega, k) &= \frac{4 \alpha}{\pi} \int_0^\infty \! \diff p \,
                        \frac{p^2}{E_f} \bigg( \frac{\omega}{v_f k} \log \frac{\omega + v_f k}{\omega - v_f k}
                                           - 1 - \frac{\omega^2 - k^2}{\omega^2 - v_f^2 k^2} \bigg)
                        \big[ n_F(E_f) + \bar{n}_F(E_f) \big]  \,.
                                    \label{eq:Pi-l}
\end{align}
In these expressions, the integrals run over fermion (electron or proton) momenta, $p$; $E_f = (p^2 + m_f^2)^{1/2}$ is the corresponding fermion energy; $v_f = p/E_f$ is the velocity; and
\begin{align}
    n_{F}(E_f) &= \frac{1}{e^{(E_f - \mu_f)/T} + 1} \, , \\
    \bar{n}_{F}(E_f) &= \frac{1}{e^{(E_f + \mu_f)/T} + 1}
\end{align}
are, respectively, the fermion and anti-fermion thermal (Fermi--Dirac) distributions at temperature $T$ and chemical potential $\mu_f$.  The differential plasmon decay rates are given by
\begin{align}
    \frac{\diff \Gamma_t(k)}{\diff E_\nu} &= \frac{G_F^2}{16 \pi^2 \alpha} Z_t(k) \frac{\omega_t(k)^2 - k^2}{\omega_t(k)}
                                   \Big[  \Big(C_V \Pi_t^{(e)}[\omega_t(k), k] - h_V \Pi_t^{(p)}[\omega_t(k), k] \Big)^2
                                               \nonumber\\
                                &\hspace{3.9cm}
                                        + \Big(C_A \Pi_A^{(e)}[\omega_t(k), k] - h_A \Pi_A^{(p)}[\omega_t(k), k] \Big)^2 \Big]
                                   \frac{k^2 + (\omega_t(k) - 2 E_\nu)^2}{4 k^3}
                                               \nonumber\\[0.2cm]
                                &\hspace{2cm}
                                        + \Big( C_V \Pi_t^{(e)}[\omega_t(k), k] - h_V \Pi_t^{(p)}[\omega_t(k), k] \Big)
                                          \Big( C_A \Pi_A^{(e)}[\omega_t(k), k] - h_A \Pi_A^{(p)}[\omega_t(k), k] \Big)
                                          \frac{\omega_t(k) - 2 E_\nu}{k^2} \,,
                                               \label{eq:Gamma-t} \\
    \frac{\diff \Gamma_l(k)}{\diff E_\nu} &= \frac{G_F^2}{32 \pi^2 \alpha} Z_l(k) \big( [\omega_l(k)]^2 - k^2 \big)^2
                                             \, \frac{\omega_l(k)}{k^5}
                                             \Big( C_V \Pi_l^{(e)}[\omega_l(k), k] - h_V \Pi_l^{(p)}[\omega_l(k), k] \Big)^2 \,
                                             \left[1 - \left(\frac{\omega_l(k) - 2 E_\nu}{k}\right)^2 \right] \,.
                                               \label{eq:Gamma-l}
\end{align}
They depend on the vector and axial-vector couplings, $C_V$ and $C_A$, between neutrinos and the electrons in the plasma (given by \cref{eq:CVe,eq:CVx,eq:CAe,eq:CAx}), and on the corresponding couplings to protons, $h_V$ and $h_A$,
\begin{align}
    h_V &= \frac{1}{2} - 2 \sin^2 \theta_W \,, \label{eq:hV} \\
    h_A &= \frac{1}{2} g_A                 \,, \label{eq:hA}
\end{align}
with the axial coupling constant $g_A \simeq 1.26$. Unlike for $C_V$ and $C_A$, the couplings to protons are the same for all neutrino flavors. The axial polarization function $\Pi_A(\omega, k)^{(f)}$ appearing in \cref{eq:Gamma-t} is
\begin{align}
    \Pi_A^{(f)}(\omega, k) &= \frac{2 \alpha}{\pi} \frac{\omega^2 - k^2}{k} \int_0^\infty \! \diff p \,
                        \frac{p^2}{E_f^2} \bigg( \frac{\omega}{2 v_f k} \log \frac{\omega + v_f k}{\omega - v_f k}
                          - \frac{\omega^2 - k^2}{\omega^2 - v_f^2 k^2} \bigg)
                        \big[ n_F(E_f) - \bar{n}_F(E_f) \big]  \,,
                                    \label{eq:Pi-A}
\end{align}
and the quantities $Z_t(k)$ and $Z_l(k)$ are the residues of the plasmon propagators at the pole, which can be interpreted as in-medium corrections to the plasmon field strengths. They are given by
\begin{align}
    Z_t(k) &\equiv \bigg[ 1 - \frac{\partial \Pi_t[\omega,k]}{\partial \omega^2} \Big|_{\omega=\omega_t(k)} \bigg]^{-1} \,,
        \label{eq:Zt} \\
    Z_l(k) &\equiv \frac{k^2}{\omega_l(k)^2} \bigg[ -\frac{\partial \Pi_l[\omega,k]}
                                                     {\partial \omega^2} \Big|_{\omega=\omega_l(k)} \bigg]^{-1} \,.
        \label{eq:Zl}
\end{align}
Here, $\Pi_t \equiv \Pi_t^{(e)} + \Pi_t^{(p)}$ and similarly for $\Pi_l \equiv \Pi_l^{(e)} + \Pi_l^{(p)}$.

It is in principle feasible to numerically compute the integrals appearing in \cref{eq:dNnu-dx-dt-dE,eq:Pi-t,eq:Pi-l,eq:Pi-A}, as well as the solutions to the dispersion relations \cref{eq:omega-t,eq:omega-l}. In practice, however, it is much more efficient to use the very accurate analytical approximations derived in Ref.~\cite{Braaten:1993jw}. Notably, the polarization functions from \cref{eq:Pi-t,eq:Pi-l,eq:Pi-A} can be approximated as
\begin{align}
    \Pi_t(\omega, k) &\simeq \omega_p^2 \, \frac{3}{2 v_*^2} \bigg( \frac{\omega^2}{k^2}
                                         - \frac{\omega^2 - v_*^2 k^2}{k^2} \frac{\omega}{2 v_* k}
                                                \log\frac{\omega + v_* k}{\omega - v_* k} \bigg) \,,
                                                        \label{eq:Pi-t-approx} \\
    \Pi_l(\omega, k) &\simeq \omega_p^2 \frac{3}{v_*^2} \bigg( \frac{\omega}{2 v_* k}
                                                          \log\frac{\omega + v_* k}{\omega - v_* k} - 1 \bigg) \,,
                                                         \label{eq:Pi-l-approx} \\
    \Pi_A(\omega, k) &\simeq \omega_A \frac{\omega^2 - k^2}{k} \frac{3}{v_*^2} \bigg(
                                 \frac{\omega}{2 v_* k} \log\frac{\omega + v_* k}{\omega - v_* k} - 1 \bigg) \,,
                                                         \label{eq:Pi-A-approx}
\end{align}
with the definitions
\begin{align}
    v_* &\equiv \frac{\omega_1}{\omega_p} \,,   \label{eq:v-star} \\
    \omega_p^2 &\equiv \frac{4 \alpha}{\pi} \int_0^\infty \! \diff p \, \frac{p^2}{E} \Big(1 - \tfrac{1}{3} v^2 \Big)
                                                              \big[ n_F(E) + \bar{n}_F(E) \big] \,, \label{eq:omega-p} \\
    \omega_1^2 &\equiv \frac{4 \alpha}{\pi} \int_0^\infty \! \diff p \, \frac{p^2}{E} \Big(\tfrac{5}{3} v^2 - v^4 \Big)
                                                              \big[ n_F(E) + \bar{n}_F(E) \big] \,, \label{eq:omega-1} \\
    \omega_A^2 &\equiv \frac{2 \alpha}{\pi} \int_0^\infty \! \diff p \, \frac{p^2}{E^2} \Big( 1 - \tfrac{2}{3} v^2 \Big)
                                                              \big[ n_F(E) - \bar{n}_F(E) \big] \,. \label{eq:omega-A}
\end{align}
$\omega_p$ is the plasma frequency, and $\omega_1$ is a measure for the first relativistic correction to it. Here, we have omitted superscripts indicating the fermion species ($f = e, p$) to shorten the notation. Plugging \cref{eq:Pi-t-approx,eq:Pi-l-approx} into \cref{eq:Zt,eq:Zl} yields approximate expressions also for $Z_t(k)$ and $Z_l(k)$. We have checked the excellent agreement between the analytic approximations and the full numerical result in the parameter region relevant for us.

\begin{figure}[hbtp]
    \centering
    \includegraphics[width=0.5\textwidth]{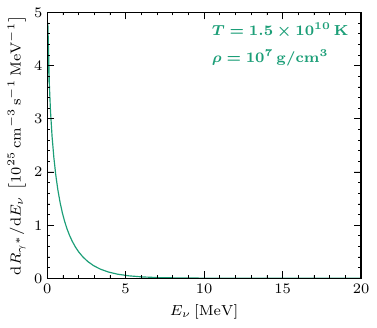}
    \caption{All-flavor neutrino spectrum due to plasmon decay, at temperatures and densities typical in a super-Eddington common-envelope event. The antineutrino spectrum is identical. Note the very large vertical scale change compared to \cref{fig:spectrum_positron}.}
    \label{fig:spectrum_plasmon}
\end{figure}

\Cref{fig:spectrum_plasmon} shows the neutrino spectrum from plasmon decay at a temperature of \SI{1.5e10}{K} and a density of \SI{e7}{g/cm^3}, characteristic of super-Eddington common-envelope events (see \cref{fig:E-loss-rate,fig:accretion_profile}). At such high temperatures, electron densities entering \cref{eq:Pi-t,eq:Pi-l,eq:Pi-A} are mainly of thermal origin, so the spectrum is not very dependent on density. As mentioned in the main text, at the high temperatures relevant for us the main production channel is positron annihilation, since plasmon decay is suppressed by $\alpha^2$. The average neutrino energy is smaller than the average plasmon energy because each plasmon decay produces a neutrino and an antineutrino. We have checked that the energy loss rate, 
\begin{equation}
\int\!\frac{4\pi k^2 \diff k}{(2\pi)^3} \, \Big[
                                   2 \omega_t(k)\, n_B[\omega_t(k)]\, \frac{\diff\Gamma_t(k)}{\diff E_\nu}
                                      + \omega_l(k)\,n_B[\omega_l(k)]\, \frac{\diff\Gamma_l(k)}{\diff E_\nu} \Big] \,,
\end{equation}
reproduces the results in Refs.~\cite{Braaten:1993jw, Itoh:1996}.

\subsection{Putting everything together}

The results in this section, together with the temperature and density profiles computed in \cref{sec:hydro}, allow us to predict the neutrino flux from a super-Eddington common-envelope event. We conservatively assume the neutron star to be completely opaque to neutrinos~\cite{Fattoyev:2017ybi}. Since neutrino emission is isotropic, this suppresses the flux emitted from a radius $r$ by a factor
\begin{equation}
    \frac{1}{2} + \sqrt{1 - \left(\frac{r_\mathrm{NS}}{r}\right)^2} \, ,
\end{equation}
an effect that reduces the neutrino flux by about 30\%. In addition, neutrinos emitted very close to the neutron star undergo gravitational redshift. Their energy measured far away, $E_\nu$, is related to the energy at production, $E_\nu^\mathrm{prod}$, by
\begin{equation}
    E_\nu = E_\nu^\mathrm{prod} \sqrt{1 - \frac{2 G M_\mathrm{NS}}{r}} \equiv E_\nu^\mathrm{prod} \sqrt{1 - \frac{r_s}{r}}\, ,
\end{equation}
with $r$ the radius where the neutrino is produced and $r_s$ the Schwarzschild radius. This effect reduces neutrino energies by about 20\%. 

Putting everything together, the neutrino flux measured at Earth is
\begin{equation}
    \frac{\diff \phi_\nu}{\diff E_\nu} = \frac{1}{4\pi d^2}\int \diff r \diff t \, 4\pi r^2 \, \left[\frac{1}{2} + \sqrt{1 - (r_\mathrm{NS}/r)^2}\right] \,\frac{\diff R}{\diff E_\nu}\left(\frac{E_\nu}{\sqrt{1-r_s/r}}, \, \rho(r, \dot{M}(t)), \, T(r, \dot{M}(t))\right) \, ,
\end{equation}
with $d$ the distance to the source and $\diff R / \diff E_\nu \equiv \diff R_{e^+e^-} / \diff E_\nu + \diff R_{\gamma^*} / \diff E_\nu$. $\rho$ and $T$ depend on the accretion rate $\dot M$, as it enters the equations of hydrodynamics via \cref{eq:continuity_2}.

\begin{figure}
    \centering
    \includegraphics[width=0.5\textwidth]{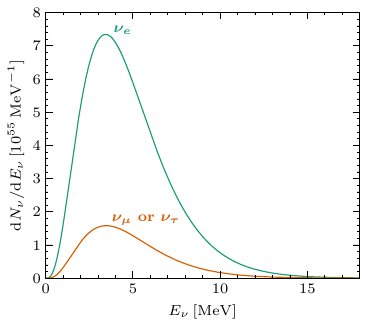}
    \caption{Neutrino spectrum from common-envelope super-Eddington accretion, using the simulation from Ref.~\cite{MacLeod:2014yda} and integrating over 3 months. Neutrino oscillations are not included. The antineutrino spectrum is identical.}
    \label{fig:spectrum_total}
\end{figure}

\Cref{fig:spectrum_total} shows the time-integrated spectrum of neutrinos produced in the last 3 months of the common-envelope simulation of Ref.~\cite{MacLeod:2014yda}. As described in the main text, most neutrinos are emitted as $\nu_e$ and $\bar{\nu}_e$ (we show the spectra before mixing effects). The average neutrino energy and flux are very similar to the order-of-magnitude estimates in the main text ($\sim \SI{4}{MeV}$ and $\sim \SI{e50}{neutrinos/s} = \SI{e57}{neutrinos/(3\,months)}$, respectively), which follow from energy-conservation arguments. 
The processes that produce neutrinos during common envelope are similar to those responsible for neutrino emission during the accretion phase of a core-collapse supernova. The three key differences are that supernova neutrinos have higher energies, the total amount of produced neutrinos is larger by a factor $\sim 10$, and emission takes place over a much shorter time.  This is because in a supernova $\dot{M} \sim \SI{0.2}{M_\odot/s}$~\cite{Janka:2012sb}, so $\langle E_\nu \rangle$ is higher by a factor $\sim 5$. The total accreted mass is also higher by a factor $\sim 50$ ($\sim \SI{2}{M_\odot}$ in a supernova and $\sim \SI{0.05}{M_\odot}$ in a super-Eddington common-envelope event), so the number of produced neutrinos is $\sim 10$ times higher.

\clearpage
\section{Neutrino oscillations}
\label{sec:neutrino_oscillations}

For completeness, we briefly mention here our treatment of neutrino oscillations (for a more complete overview of oscillations in a dense environment, see Ref.~\cite{Dighe:1999bi}). We include adiabatic transitions, as non-adiabatic corrections only modify our results by $\lesssim 10\%$~\cite{Kuo:1989qe, Maltoni:2015kca}. We adopt the mixing parameters from Ref.~\cite{Esteban:2020cvm} (uncertainties on these parameters, including the mass ordering, will be strongly reduced in the near future~\cite{JUNO:2024jaw, DUNE:2020jqi, Hyper-Kamiokande:2018ofw}). Overall, oscillations change the common-envelope signal by a factor of few.

Thermal neutrino emission produces the same amount of $\nu_\mu$ and $\nu_\tau$. Furthermore, at the energies we consider it is very challenging to experimentally distinguish between them~\cite{Brdar:2023ttb}. Hence, we can consider an effective 2-flavor problem, and write the $\nu_e$ flux at Earth as
\begin{equation}
    \phi_e = \phi_e^0 P_{ee} + \phi_x^0 (1-P_{ee}) \, ,
\end{equation}
where $\phi_e^0$ and $\phi_x^0$ are the initial fluxes of $\nu_e$ and $\nu_\mu$ (or $\nu_\tau$), respectively; $P_{ee}$ is the $\nu_e \rightarrow \nu_e$ transition probability. The average of the $\nu_\mu$ and $\nu_\tau$ fluxes at Earth is given by
\begin{equation}
    \phi_x \equiv \frac{\phi_\mu + \phi_\tau}{2} = \frac{1}{2}\left[ \phi_e^0 (1- P_{ee}) + \phi_x^0(1+P_{ee})\right] \, .
\end{equation}
Analogously, for antineutrinos
\begin{align}
    \phi_{\bar e} & = \phi_{\bar e}^0 P_{\bar{e}\bar{e}} + \phi_{\bar{x}}^0 (1-P_{\bar{e}\bar{e}}) \, , \\
    \phi_{\bar x} \equiv \frac{\phi_{\bar\mu} + \phi_{\bar\tau}}{2} & = \frac{1}{2}\left[ \phi_e^0 (1- P_{\bar{e}\bar{e}}) + \phi_x^0(1+P_{\bar{e}\bar{e}})\right] \, ,
\end{align}
where bars indicate the corresponding antineutrino flux and $P_{\bar{e}\bar{e}}$ is the $\bar{\nu}_e \rightarrow \bar{\nu}_e$ transition probability.

In a dense environment, transitions happen mainly when the density equals the resonant density, $\sim 10^3$--$\qty{e4}{g/cm^3}$ for the atmospheric mass splitting and $\sim 10$--$\SI{30}{g/cm^3}$ for the solar mass splitting. Neutrinos from a super-Eddington common-envelope event cross both resonance regions (see \cref{fig:accretion_profile}), but which resonances they actually experience depends on the neutrino mass ordering. 

In the normal mass ordering (NO), an initial $\nu_e$ experiences both resonances, exiting the star as a $\nu_3$ mass eigenstate, and $P_{ee} = |U_{e3}|^2 \simeq 0.022$. In the inverted mass ordering (IO), only the solar resonance lies in the neutrino sector, and $P_{ee} = |U_{e2}|^2 \simeq 0.30$. For antineutrinos, the picture is different: in the NO, all resonances lie in the neutrino sector, and $P_{\bar{e}\bar{e}} = |U_{e1}|^2 \simeq 0.68$; in the IO, the atmospheric resonance lies in the antineutrino sector, and $P_{\bar{e}\bar{e}} = |U_{e3}|^2 \simeq 0.022$. This is why for the IO the signal in Super-K and JUNO, which rely on inverse beta decay, is suppressed in \cref{fig:rates}. The remaining events in these experiments are then mostly due to $\bar\nu_\mu$ and $\bar\nu_\tau$ converting into $\bar\nu_e$. Since the number of initially produced $\bar\nu_\mu$ and $\bar\nu_\tau$ is smaller than the number of initially produced $\bar\nu_e$, the overall signal is suppressed. Neutrino--electron scattering in DUNE has some sensitivity to all neutrino flavors, so the expected rates are more similar for both mass orderings.

\section{Details of experiment simulations}
\label{sec:experiments}

We provide details on how we simulate the various experiments considered in \cref{fig:rates,fig:reach} in the main text. We aim to be conservative in our assumptions, leaving a lot of room for improvement by dedicated analyses carried out within the experimental collaborations.

In each case we consider, the neutrino energy is estimated by reconstructing the energy of the electron or positron produced when the neutrino interacts. The number of events with reconstructed electron/positron energy $\in [E_e^\mathrm{min}, E_e^\mathrm{max}]$ from a common-envelope system is given by
\begin{equation}
    N_\mathrm{evt} = \int_{E_e^\mathrm{min}}^{E_e^\mathrm{max}} \diff E_e^\mathrm{rec}
                     \int_0^\infty \diff E_e
                     \int_0^\infty \diff E_\nu \, \frac{\diff \phi_\nu(E_\nu)}{\diff E_\nu} \,
                         \frac{\diff \sigma(E_\nu, E_e)}{\diff E_e}
                         \, N_\mathrm{targets} \, f_\mathrm{rec}(E_e^\mathrm{rec}, E_e) \,
                         \epsilon(E_e) \, ,
\end{equation}
with $E_e^\mathrm{rec}$ the reconstructed electron/positron energy, $E_e$ the true electron/positron energy, $E_\nu$ the neutrino energy, $\diff \sigma / \diff E_e$ the differential neutrino interaction cross section, $N_\mathrm{targets}$ the number of targets in the detector, $f_\mathrm{rec}(E_e^\mathrm{rec}, E_e)$ the energy response function (i.e., the differential probability for an electron/positron with true energy $E_e$ to be reconstructed with energy $E_e^\mathrm{rec}$), and $\epsilon$ the detection efficiency.

We compute the statistical significance of the signal against a background-only hypothesis with a Poissonian likelihood ratio,
\begin{equation}
    \Delta \chi^2 = 2 \sum_\mathrm{bins} \bigg( N_\mathrm{bkg} - N_\mathrm{total} + N_\mathrm{total} \log \frac{N_\mathrm{total}}{N_\mathrm{bkg}} \bigg) \, ,
\end{equation}
with $N_\mathrm{bkg}$ the number of background events in a given bin, and $N_\mathrm{total} \equiv N_\mathrm{evt} + N_\mathrm{bkg}$. For \cref{fig:rates}, we use \SI{2}{MeV}-wide energy bins. For \cref{fig:reach}, we specify the bin widths below.

\subsection{Super-K (archival)}

Our simulation for archival Super-Kamiokande data closely follows Ref.~\cite{Super-Kamiokande:2021jaq}, which looked for inverse beta decay events due to diffuse supernova background neutrino interactions by searching for positrons together with delayed gamma rays due to neutron capture on protons.

We take the inverse beta decay cross section at order $E_\nu/m_N$, with $m_N$ the nucleon mass, from Refs.~\cite{Vogel:1999zy, Strumia:2003zx}. The fiducial volume contains \SI{22.5}{ktonne} of water, corresponding to $N_\mathrm{targets} \simeq 1.5\times 10^{33}$ free protons. For energy reconstruction, we assume Gaussian smearing with the Super-Kamiokande low-energy resolution~\cite{Super-Kamiokande:2016yck}
\begin{equation}
    \sigma(E_e)/\mathrm{MeV} = -0.0839 + 0.349 \sqrt{E_e/\mathrm{MeV}} + 0.0397 \, (E_e/\mathrm{MeV}) \, .
\end{equation}
We extract the detection efficiency, rising from $\sim 5\%$ at \SI{8}{MeV} to $\sim 20\%$ at \SI{20}{MeV}, from Fig.~18 in Ref.~\cite{Super-Kamiokande:2021jaq}. At the energies we consider, backgrounds are mainly due to accidental coincidences, $^9$Li produced via spallation and undergoing beta decay together with neutron emission (which leaves the same signature as inverse beta decay), neutral current interactions of atmospheric neutrinos (which induce nuclear gamma-ray emission), and a small component from reactor antineutrino and charged current atmospheric neutrino interactions. We extract the backgrounds from Fig.~24 in Ref.~\cite{Super-Kamiokande:2021jaq}, rescaling them to the 3-month exposure we consider. We consider positron kinetic energies $E^\mathrm{kin}_e \equiv E_e - m_e c^2$ above \SI{7.5}{MeV}, and 2-MeV bins~\cite{Super-Kamiokande:2021jaq}.

\subsection{Super-K (low Gd)}

Our simulation for the first Super-Kamiokande run with gadolinium loading~\cite{Beacom:2003nk} (corresponding to 0.01\% mass concentration of gadolinium) closely follows Ref.~\cite{Super-Kamiokande:2023xup}. The cross section and energy reconstruction are the same as for \emph{Super-K (archival)}. We extract the detection efficiency, rising from $\sim 15\%$ at \SI{8}{MeV} to $\sim 25\%$ at \SI{20}{MeV}, from Fig.~1 in Ref.~\cite{Super-Kamiokande:2023xup}. The main background sources are the same as for \emph{Super-K (archival)}. Since some bins in Ref.~\cite{Super-Kamiokande:2023xup} are wider than \SI{2}{MeV}, we also simulate the backgrounds, finding good agreement with Ref.~\cite{Super-Kamiokande:2023xup}. In particular, we neglect the subleading accidental coincidence contribution~\cite{Super-Kamiokande:2023xup}, we compute theoretically the $^9$Li decay spectrum and rescale it to the normalization in Fig.~2 in Ref.~\cite{Super-Kamiokande:2023xup} (rescaling also to the 3-month exposure we consider), we extract the atmospheric contribution from Fig.~2 in Ref.~\cite{Super-Kamiokande:2023xup} (rescaling to the 3-month exposure we consider), and we simulate the reactor background using the publicly available \texttt{SKReact} code. We consider positron kinetic energies above \SI{7.5}{MeV}, and 2-MeV bins~\cite{Super-Kamiokande:2023xup}.

\subsection{Super-K (Gd, optim.)}

For an optimistic future upgrade of Super-Kamiokande, we assume a gadolinium concentration large enough to increase the detection efficiency for inverse beta decay events to $75\%$. This efficiency corresponds to the limit in which all neutrons are captured on gadolinium~\cite{Super-Kamiokande:2023xup}. We further assume that background levels can be kept as in \emph{Super-K (Gd)}, with reactor backgrounds increased proportionally to the detection efficiency increase. We take the same cross section and energy resolution as in \emph{Super-K (archival)}. We consider positron kinetic energies above \SI{3.5}{MeV} (at the sensitivity limit shown in \cref{fig:reach}, backgrounds overwhelm the signal below $\sim \SI{7}{MeV}$), and 0.5-MeV bins.

\subsection{JUNO}

Our JUNO simulation closely follows Refs.~\cite{JUNO:2021vlw, JUNO:2023vyz}.
The cross section is the same as for \emph{Super-K (archival)}. The detector contains $1.44\times 10^{33}$ free protons~\cite{JUNO:2015zny}. For the energy resolution, we assume Gaussian smearing with~\cite{JUNO:2021vlw}
\begin{equation}
    \sigma(E_\mathrm{vis}) = E_\mathrm{vis} \sqrt{a^2/E_\mathrm{vis} + b^2 + c^2/E_\mathrm{vis}^2} \, ,
\end{equation}
where $E_\mathrm{vis} = E_e + m_e c^2$ the deposited energy after positron annihilation, $a=\SI{0.0261}{\sqrt{MeV}}$, $b=0.0082$, and $c=\SI{0.0123}{MeV}$. We assume 73\% detection efficiency~\cite{JUNO:2021vlw}. At the energies we consider, backgrounds are mainly due to reactor neutrinos, $^9$Li and $^8$He produced via spallation (which can undergo beta decay together with neutron emission), and neutral current interactions of atmospheric neutrinos. We extract the background from Fig.~9 in Ref.~\cite{JUNO:2021vlw} and Fig.~6 in Ref.~\cite{JUNO:2023vyz}, rescaling it to the 3-month exposure we consider. We consider $E_\mathrm{vis} > \SI{5}{MeV}$ (at the sensitivity limit shown in \cref{fig:reach}, backgrounds overwhelm the signal below $\sim \SI{7.5}{MeV}$), and 0.5-MeV bins.

\subsection{Hyper-K}

To simulate the Hyper-Kamiokande detector, we rescale signals and backgrounds from \emph{Super-K (archival)} by a factor 187/22.5, reflecting the increase in fiducial volume of the 1-tank design. We conservatively assume the same energy resolution as in Super-K~\cite{Hyper-Kamiokande:2018ofw}. We also increase $^9$Li spallation backgrounds by a factor of 2.7 due to the detector's shallower depth~\cite{Hyper-Kamiokande:2018ofw}. We consider positron kinetic energies above \SI{7.5}{MeV}, and 2-MeV bins~\cite{Super-Kamiokande:2021jaq}.

\subsection{Hyper-K (Gd)}

To simulate the Hyper-Kamiokande detector with gadolinium loading, we rescale signals and backgrounds from \emph{Super-K (Gd, optim.)} by a factor 187/22.5, reflecting the increase in fiducial volume of the 1-tank design. We conservatively assume the same energy resolution as in Super-K~\cite{Hyper-Kamiokande:2018ofw}. We also increase $^9$Li spallation backgrounds by a factor of 2.7 due to the detector's shallower depth~\cite{Hyper-Kamiokande:2018ofw}. We consider positron kinetic energies above $\SI{3.5}{MeV}$ (at the sensitivity limit shown in \cref{fig:reach}, backgrounds overwhelm the signal below $\sim \SI{8}{MeV}$), and 0.5-MeV bins.

\subsection{DUNE}

For our DUNE simulation, we closely follow Refs.~\cite{Capozzi:2018dat, Zhu:2018rwc, DUNE:2020ypp}, which analyzed the potential of the experiment to detect solar neutrinos.

We take the neutrino--electron scattering cross section for the different neutrino flavors from Ref.~\cite{Formaggio:2012cpf}. We assume a \SI{40}{ktonne} fiducial volume, corresponding to the four-module design. This corresponds to $N_\mathrm{targets}\simeq 1.1\times 10^{34}$ electrons. We assume Gaussian smearing with 7\% energy resolution~\cite{Castiglioni:2020tsu} in electron kinetic energy. We extract the detection efficiency from Fig.~7.17 in Ref.~\cite{DUNE:2020ypp}. At the energies we consider, backgrounds are mainly due to neutrons from the surrounding rock and spallation products. We take them from Refs.~\cite{Capozzi:2018dat, Zhu:2018rwc}, rescale them to the 3-month exposure we consider, and reduce them by considering a forward cone of half-angle $40^\circ$~\cite{Capozzi:2018dat, Zhu:2018rwc}. We consider electron kinetic energies above \SI{5}{MeV} (at the sensitivity limit shown in \cref{fig:reach}, backgrounds overwhelm the signal below $\sim \SI{7.5}{MeV}$), and 0.5-MeV bins.

As discussed in the main text, in \cref{fig:rates} we assume that solar neutrino interactions on ${}^{40}$Ar can be efficiently removed by identifying final-state gamma rays from nuclear deexcitation. In \cref{fig:reach} in the main text, we also show the distance reach if gamma rays can not be identified. In that case, there are additional backgrounds from interactions of solar and common-envelope neutrinos with ${}^{40}$Ar. We include them by taking the interaction cross section from the public code \texttt{SNoWGLoBES}, and reduce them by considering a forward cone of half-angle $40^\circ$~\cite{Capozzi:2018dat, Zhu:2018rwc}.

\subsection{IceCube}

The IceCube experiment can detect core-collapse supernova neutrino bursts by observing a collective rise in the average count rate across all optical modules~\cite{IceCube:2011cwc}. Hence, in principle it might be sensitive to common-envelope neutrinos too.

However, as described at the end of \cref{sec:neutrino_emission}, a common-envelope signal has $\sim 10$ times less neutrinos than a supernova signal, and it is spread over several months instead of $\sim 10$ seconds. Hence, the neutrino event rate is many orders of magnitude smaller for a common-envelope signal than for a supernova signal. Following Ref.~\cite{IceCube:2011cwc}, we estimate that IceCube is only sensitive to common-envelope events within $\sim 100\,\mathrm{pc}$ of Earth. Since this is below the smallest distance we consider in \cref{fig:reach}, we do not include IceCube in our main results.

\section{Distance reach for the inverted mass ordering}
\label{sec:distance}

\begin{figure}[hbtp]
    \centering
    \includegraphics[width=0.5\textwidth]{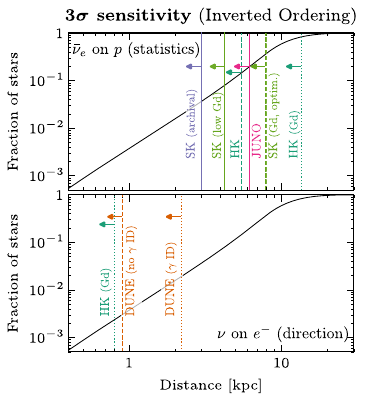}
    \caption{Same as \cref{fig:reach}, but assuming the inverted mass ordering.}
    \label{fig:reach_IO}
\end{figure}

In \cref{fig:reach} in the main text, we show the distance reach to super-Eddington neutron-star common-envelope events assuming the normal mass ordering, which is currently favored over the inverted mass ordering by $\sim 2.5\sigma$~\cite{Esteban:2020cvm}. \Cref{fig:reach_IO} shows the distance reach for the inverted ordering. As described in the main text, the event rate gets reduced for the inverse beta decay channel and increased for neutrino--electron scattering. Nevertheless, the fraction of Milky Way stars covered only changes by at most a factor of few. 

\end{document}